\documentclass[showpacs,preprintnumbers,amsmath,amssymb,10pt]{revtex4}

\usepackage{psfig}
\usepackage{graphicx}
\usepackage{dcolumn}
\usepackage{bm}

\newcommand{\s}{\ensuremath{\psi(t,r)}}
\newcommand{\n}{\ensuremath{\nu(t,r)}}

\newcommand{\e}{equation\;} 
\newcommand{\M}{\ensuremath{{\cal M}}}

\newcommand{\X}{\ensuremath{{\cal X}}}

\begin{document}

\preprint{}
\title{Gravitational collapse of an isentropic perfect fluid with a 
linear equation of state}
\author{Rituparno Goswami}
\email{goswami@tifr.res.in}
\author{Pankaj S Joshi}
\email{psj@tifr.res.in}
\affiliation{ Department of Astronomy and Astrophysics\\ Tata
Institute of Fundamental Research\\ Homi Bhabha Road,
Mumbai 400 005, India}

\begin{abstract} We investigate here the gravitational 
collapse end states for a spherically symmetric perfect fluid with an 
equation of state $p=k\rho$. It is shown that given a regular initial 
data in terms of the density and pressure profiles at the initial epoch 
from which the collapse develops, the black hole or naked 
singularity outcomes depend on the choice of rest of the free functions 
available, such as the velocities of the collapsing shells, and 
the dynamical evolutions as allowed by Einstein equations. This 
clarifies the role that equation of state and initial data play 
towards determining the final fate of gravitational collapse.   

\end{abstract}

\maketitle

\section{Introduction}
Considerable interest is seen in recent years to 
examine the final fate of gravitational collapse of a massive 
matter cloud within the framework 
of Einstein's theory of gravity. This is due to the importance 
of this problem in black hole physics, 
and the related cosmic censorship conjecture which 
states the spacetime singularities of collapse must be hidden 
within black holes, and would not communicate with faraway observers 
in the spacetime (for some recent reviews, see e.g.
[1-5].
The assumption that continual collapse of a 
matter cloud, such as a massive star which has exhausted its nuclear
fuel, ends only in a black hole (BH) and not naked singularity (NS)
is crucial to many of the considerations in the physics
and astrophysics of black holes.
Several dynamical collapse scenarios have been 
investigated extensively from such a perspective, which include
the radiation collapse within the framework of a Vaidya metric 
(see e.g. 
\cite{book1}, 
and references therein), collapse of a dust cloud  
[6-10],
some considerations on perfect fluids (many of these being numerical)
[11-19],
massless scalar fields
[20-26],
and also more general matter fields
[27-29].

Our purpose in this note is to study 
analytic models of spherical gravitational collapse of a perfect fluid 
with an equation of state $p=k\rho$. This is of interest because 
this is a well-studied and extensively used case in astrophysics, 
which offers a physically interesting model. 
While modeling a realistic collapse, one has to consider the 
{\it equation of state (EOS)} of the collapsing matter as an additional 
constraint over the Einstein equations. This is a long standing
interesting question in the arena of dynamical collapse theories    
as to how a physically realistic EOS affects the evolution of a collapse 
in terms of it's final state.
In the later stages of collapse one could not ignore the 
pressures, and hence the case when matter is an isentropic perfect 
fluid with a linear equation of state offers a useful scenario 
to examine collapse in ultra-relativistic limits.

We show here that a perfect fluid collapse could end 
in either of the BH/NS final states, depending on the nature of
the initial data, and the 
allowed evolutions for the collapsing matter. 
Given a regular initial data for matter in terms of the regular 
density and pressure profiles, the exact outcome in terms of the 
above depends on the choice of rest of the free functions available, 
such as the velocities of the collapsing shells, and the
allowed evolutions for the collapsing matter.
Our results thus provide some insight on the role that
equation of state and initial data play to determine the final fate
of continual gravitational collapse in terms of the BH/NS
end states.

The basic regularity conditions and the Einstein equations are
given in Section 2. The collapse is examined in Section 3, and in 
Section 4 we study how the nature of the singularity is determined by
the initial data and the evolution chosen. Some conclusions are 
summarized in Section 5.

\section{Einstein equations and regularity conditions}

The spacetime geometry within the spherically symmetric collapsing 
cloud can be described by the metric in the comoving   
coordinates $(t,r,\theta,\phi)$ as given by,
\begin{equation}
ds^2=-e^{2\n}dt^2+e^{2\s}dr^2+R^2(t,r)d\Omega^2
\label{eq:metric}
\end{equation}
where $d\Omega^2$ is the line element on a two-sphere. 
In this reference frame the energy-momentum tensor for a perfect fluid 
is given by,
\begin{equation}  
T^{ij}=\left(\rho+p\right)V^iV^j+pg^{ij}
\label{eq:emtensor}
\end{equation}
where $V^i$ is unit timelike vector. We also take the matter 
fields to satisfy the {\it weak energy condition}, i.e. the energy density 
measured by any local observer is non-negative. Then we must have,
\begin{equation}
T_{ik}V^iV^k\ge0
\end{equation}
which amounts to,
\begin{equation}
\rho\ge0;\; \rho+p\ge0;
\end{equation} 
Since we consider here an {\it isentropic} fluid, whose pressure is a 
linear function of the density only, the equation of state of the collapsing
matter is given by,
\begin{equation}
p(t,r)=k\rho(t,r)
\label{eq:pf}
\end{equation}
where $k\in(-1,1]$ is a constant. We note that the cases of negative $k$
are the {\it dark energy fluids} with negative pressures.
Now for the metric (\ref{eq:metric}) the Einstein equations 
take the form (in the units $8\pi G=c=1)$
\begin{equation}
\rho=\frac{F'}{R^2R'}=-\frac{1}{k}\frac{\dot{F}}{R^2\dot{R}}
\label{eq:ein1}
\end{equation}
\begin{equation}
\nu'=-\frac{k}{k+1}\left[\ln(\rho)\right]'
\label{eq:ein2}
\end{equation}
\begin{equation}
R'\dot{G}-2\dot{R}\nu'G=0
\label{eq:ein3}
\end{equation}
\begin{equation}
G-H=1-\frac{F}{R}
\label{eq:ein4}
\end{equation}
where,
\begin{eqnarray}
G(t,r)=e^{-2\psi}(R')^2; && H(t,r)=e^{-2\nu}(\dot{R})^2
\label{eq:ein5}
\end{eqnarray}
The arbitrary function $F=F(t,r)$ here has an 
interpretation of the mass function for the cloud, and it
gives the total mass in a 
shell of comoving radius 
$r$ on any spacelike slice $t=const$. We have $F\ge0$ from the energy 
conditions. In order to 
preserve the regularity at the initial epoch, we have  
$F(t_i,0)=0$, that is, the mass function should vanish at the center 
of the cloud. Since we are considering collapse, we have $\dot R<0$,
i.e. the physical radius $R$ of the cloud keeps decreasing 
in time and
ultimately reaches $R=0$, which denotes the singularity where all
matter shells collapse to a zero physical radius.
We can use the scaling freedom available for the radial co-ordinate 
$r$ to write $R=r$ at the initial epoch $t=t_i$. 
Let us introduce a function $v(t,r)$ as defined by,
\begin{equation}
v(t,r)\equiv R/r 
\label{eq:R}
\end{equation}
we then have $R(t,r)=rv(t,r)$, and 
\begin{eqnarray}
v(t_i,r)=1; & v(t_s(r),r)=0; & \dot{v}<0
\label{eq:v}
\end{eqnarray}
The time $t=t_s(r)$ corresponds to the shell-focusing 
singularity at $R=0$, where all the matter shells collapse to a 
vanishing physical radius.

The description of the singularity in terms of $v(t,r)$ has
the following advantage. The physical radius goes to the value
zero at the shell-focusing singularity, but we also have $R=0$
at the regular center of the cloud at $r=0$. This is to be 
distinguished from the genuine singularity by the fact, for 
example, that the density and other physical quantities including
the curvature scalars are all finite at a regular center $r=0$,
even though $R=0$ holds there. This is achieved, as we point out below,
by a suitable behaviour of the mass function, which should go to 
a vanishing value sufficiently fast in the limit of approach to
the regular center where even though $R$ goes to zero
the density must remain finite. On the other hand, we note that
at $t=t_i$ we have $v=1$, and then as the collapse evolves $v$ 
continuously decreases to become zero 
only at the singularity, i.e. $v=0$ uniquely corresponds to the 
genuine spacetime singularity at $R=0$.

We thus see that there are now five total field equations 
with five unknowns, which are $\rho$, $\psi$, $\nu$, $R$, and $F$.
Solutions of these 
equations, subject to the weak energy condition and the given 
regular initial data for collapse at the initial spacelike 
surface $t=t_i$, determine the matter distribution and 
metric of the space-time. We then have specific time evolutions 
of the initial data which define the collapse final states. 
It turns out that there exist classes of solutions, which give
either a black hole or a naked singularity
as the end state of the collapse, depending on the nature of
the initial data, and the class of evolutions chosen.

\section{Collapsing Matter Clouds}

It is now possible to consider the gravitational collapse 
of a perfect fluid within this framework as we discuss below. 
The regularity conditions as defined above
set up the initial data at the initial surface $t=t_i$ from
which the collapse develops. We now consider the collapse equations 
which allow us to see when the singularity occurs, and how the 
initial data and the classes of evolutions as governed by the 
Einstein equations lead to the formation of the spacetime 
singularity.

We consider a general mass function $F(t,r)$ for the
collapsing cloud, which can be written as
\begin{equation}
F(t,r)=r^3\M (r,v)
\label{eq:mass}
\end{equation}
where $\M$ is a regular and suitably differentiable function, 
and $\M>0$. 
As seen from equations below, the regularity and finiteness 
of the density profile at the initial epoch $t=t_i$, and at all other 
regular epochs before the singularity at $R=0$ develops,
requires that $F$ goes as $r^3$
close to the center.
Hence we note that since $\M$ is a general (at least $C^2$) 
function, 
the equation \e(\ref{eq:mass}) is not really any ansatz, or a specific 
choice, but quite a generic form of the mass profile
for the collapsing cloud.
We note that 
[19] 
considered mass profiles which are analytic (with $F(t,r)=F(r,R)$), 
and perfect fluids
could form naked singularity, but we do not impose such an assumption 
on mass function here.

Then equation (\ref{eq:ein1}) gives,
\begin{equation}  
\rho=\frac{3\M+r\left[\M_{,r}+\M_{,v}v'\right]}{v^2(v+rv')}=   
-\frac{1}{k}\frac{\M_{,v}}{v^2}
\label{eq:rho}
\end{equation}
The regular density distribution at the initial epoch is given by,
\begin{equation}
\rho_0(r)=3\M(r,1)+r\M(r,1)_{,r}
\label{eq:rho0}
\end{equation}
It is evident that, in general, as $v\rightarrow 0$, $\rho\rightarrow\infty$.
That is, both the density and pressure blow up at the singularity. We note
from equation (\ref{eq:rho}) that $\rho=\rho(r,v)$ and hence $v'=f(r,v)$.
Now rewriting  equation (\ref{eq:rho}) we get,
\begin{equation}
3k\M + kr\M_{,r} + Q(r,v)\M_{,v}=0  
\label{eq:eos}  
\end{equation}
where,
\begin{equation}
Q(r,v)=(k+1)rv'+v
\label{eq:q}
\end{equation}  
Now the above equation (\ref{eq:eos})
has a general solution of the form
\cite{diff},  

\begin{equation}
{\cal F}(X,Y)=0
\label{eq:solution}
\end{equation}  
where $X(r,v,\M)$ and $Y(r,v,\M)$ are the solutions
of the system of equations,
\begin{equation}
\frac{d\M}{3k}=\frac{dr}{kr}= \frac{dv}{Q}
\label{eq:auxilliary}
\end{equation}   
Amongst all the classes of solutions of $\M(r,v)$ as given by
equation (\ref{eq:solution}),
only those are to be considered which obey the energy condition,
which are regular,
and which in the limit of $v\rightarrow 0$ give
$\rho\rightarrow\infty$.
In other words, the equation of state as given by the perfect fluid
condition $p=k\rho$, and the energy condition isolate the
class of the mass functions to be considered.

We can now directly integrate equation (\ref{eq:ein2}) to get,
\begin{equation}
\nu(r,v)=-\frac{k}{k+1}\left[\ln(\rho)\right] 
\label{eq:nu}
\end{equation}
Let us now define a suitably differentiable function $A(r,v)$ in the
following way,
\begin{equation}
\nu'(r,v)=A(r,v)_{,v}R'
\label{eq:A}
\end{equation}
That is $A(r,v)_{,v} \equiv \nu'/R'$. 
We note that equation (14) can in principle give solutions which 
are not regular, or such that the function $A(r,v)$ as defined above 
is not regular. However, regularity of $\M$ and $A$ is a necessary
assumption we make here, which will be used later also (see e.g. 
remarks after equation (30)). Hence, only the class of regular 
solutions is to be considered.

Our main interest
here is in studying the shell-focusing singularity at $R=0$ which is 
physical singularity where all the shells collapse to zero radius.
Hence we assume that there are no shell-crossing singularities 
in the spacetime, where $R'=0$ and so the function $A(r,v)$ is
well-defined.

Some comments are in order here on our assumption that $R'>0$,
that is, we consider the situation with no shell-crossing singularities.
This is because, it is generally believed (see e.g.
\cite{shell}
that such singularities can be possibly removed from the spacetime 
as they are typically gravitationally weak, and also because 
spacetime extensions have been constructed through the 
same in certain cases. Under the situation, we are interested
only in examining the nature of the shell-focusing singularities
at $R=0$, which are genuine curvature singularities, where the
physical radii for all collapsing shells vanish, and the
spacetime necessarily terminates without extension.

Specifically, $R'>0$ implies that we must have $v + rv'> 0$. 
Since $v$ is necessarily positive, it follow that this will be
satisfied whenever $v'$ is greater or equal to zero, or even
when it is negative the magnitude of $rv'$ should be less then
that of $v$. Later in this section we shall derive an expression for  
the quantity $v'$, in terms of the initial data and the other
free evolutions as allowed by the Einstein equations. Hence it
follows that we can specifically state the condition for
avoidance of shell-crossings in terms of the behaviour of
these functions. In particular, it turns out that whenever the
singularity curve $t_s(r)$ (which corresponds to $R=0$) is
increasing (or when it decreases at a sufficiently slow rate) 
with a slope greater or equal to zero at the origin,
the shell-crossing singularities are avoided at least in
the vicinity of the regular center $r=0$. We then have a ball
of finite radius around the regular center which contains no
shell-crossings till the final singularity formation at
$R=0$. We shall, however, not go into further details here.

At the initial epoch we have,
\begin{equation}
\left. A(r,v)_{,v}\right|_{v=1}=-\frac{k}{k+1}\left
[\frac{\rho_0'(r)}{\rho_0(r)}\right]
\label{eq:A1}
\end{equation}
In fact, for all epochs, the relation between the function $\M$ and $A$
is given by equation (\ref{eq:ein2}) as $A_{,v}R'=-\frac{k}{k+1}\ln
\left[-\frac{\M_{,v}}{kv^2}\right]'$.
If we consider a smooth initial profile, {\it i.e} the gradient 
of the initial density vanishes at the center, then we must have 
$A(r,v)=rg(r,v)$, where $g(r,v)$ is another suitably differentiable function.

Now using equation (\ref{eq:A}) we can integrate (\ref{eq:ein3}) to get,
\begin{equation}
G(r,v)=b(r)e^{2rA}
\label{eq:G}
\end{equation}
Here $b(r)$ is another arbitrary function of the comoving coordinate
$r$. A comparison
with dust collapse models interprets $b(r)$ as the velocity function for
the collapsing shells. Following this parallel, we 
can write,
\begin{equation}
b(r)=1+r^2b_0(r)
\label{eq:veldist}
\end{equation}
Finally, using equations (\ref{eq:G}), (\ref{eq:nu}) and 
(\ref{eq:veldist}) in
(\ref{eq:ein4}) we have,
\begin{equation}
\sqrt{v}\dot{v}=-\rho^{-\frac{k}{k+1}}\sqrt{e^{2rA}vb_0(r)+vh(r,v)+\M(r,v)}
\label{eq:collapse1}
\end{equation}
where,
\begin{equation}
h(r,v)=\frac{e^{2rA}-1}{r^2}
\label{eq:h}
\end{equation}  
Integrating the above equation we have,
\begin{equation}
t(v,r)=\int_v^1\frac{\sqrt{v}dv}{\rho^{-\frac{k}{k+1}}\sqrt{e^{2rA}vb_0+vh+\M}}
\label{eq:scurve1}
\end{equation}
Note that the variable $r$ is treated as a constant in the above equation. 
Close to the center we can write $t(v,r)$ as,
\begin{equation} 
t(v,r)=t(v,0)+r\X(v)+{\cal O}(r^2)
\label{eq:scurve2}
\end{equation}
Here the function $\X(v)$ is given by,
\begin{equation}
\X(v)=-\frac{1}{2}\int_v^1dv\frac{\sqrt{v}B_1(0,v)}{B(0,v)^{\frac{3}{2}}}
\label{eq:tangent1}
\end{equation}
where,
\begin{equation}
B(r,v)=\rho^{-\frac{k}{k+1}}\sqrt{e^{2rA}vb_0+vh+\M};\;B_1=B_{,r}
\label{eq:B}
\end{equation}

In order to obtain equation (28), we note that we require  
the integral (27) could be differentiated. This is possible because it is 
finite by definition, and then we need all the functions, 
namely $A(r,v), b_0(r)$ and $M(r,v)$ to be suitably differentiable.
In our case we require them to be at least $C^2$ for $r$ not equal 
to zero, and $C^1$ for $r=0$.

Thus we see that the time taken for the central shell to 
reach the singularity is given as
\begin{equation}
t_{s_0}=\int_0^1\frac{\sqrt{v}dv}{B(0,v)}
\label{eq:scurve3}
\end{equation}
The time for other shells to reach the singularity  
is given by the following, which defines the singularity
curve developing in the spacetime as end result of collapse,
\begin{equation}
t_s(r)=t_{s_0}+r\X(0)+{\cal O}(r^2)
\label{eq:scurve4}
\end{equation}
It is now clear that the value of the quantity $\X(0)$, which
represents the tangent to the singularity curve, depends 
on the functions $b_0$, $\M$ and $h$, which have the initial
values as dictated by the initial data at $t=t_i$, and which are
functions of $r$ and $v$ as the case may be. Hence, a 
given set of density and velocity 
profiles, together with the evolutions 
chosen, completely determines the tangent at the center to the 
singularity curve.
Further, from \e(\ref{eq:collapse1}), we get, for a constant $v$ surface,
\begin{equation}
\sqrt{v}v'=\X(v)B(0,v)+{\cal O}(r)
\label{eq:scurve5}
\end{equation}

We note that as seen above, $\X(0)$ involves functions depending 
on the initial data, and also the evolutions $A$ and $\M$. 
The relation between the functions $A$ and $\M$ is given by the perfect
fluid equation of state, for which we have shown that solutions exist,
and the Einstein equations as noted earlier.
Among different classes of solutions only those are to be considered which
ensure the density and pressures to blow up at the singularity.

One has now to understand the structure of this singularity,
and to examine when it will be visible, and when covered within an
event horizon of gravity, i.e. hidden within a black hole.

\section{Nature of the singularity}

It is now possible to see for a perfect fluid collapse 
with a linear equation of state, how the initial data and the allowed 
evolutions determine 
the final fate of collapse in terms of either a black hole or 
a naked singularity. The apparent horizon 
within the collapsing cloud is given by $R=F$. If the neighborhood 
of the center gets trapped earlier than the singularity, then it 
is covered and a black hole results, otherwise it is visible with 
non-spacelike future directed trajectories escaping from it.
In other words, we examine below when there will be families of
null geodesics existing, which will be future directed and
outgoing, and which terminate in the past at the singularity,
thus making the communication from the singularity to an outside
observer possible, as opposed to a black hole situation where
this will not be the case.

In order to consider the possible existence of such trajectories and 
to examine the nature of the central singularity at $R=0$, $r=0$, 
let us consider the equation for outgoing radial null geodesics which
is given by,
\begin{equation}
\frac{dt}{dr}=e^{\psi-\nu}
\label{eq:null1}
\end{equation}
The singularity occurs at $v(t_s(r),r)=0$, i.e. $R(t_s(r),r)=0$. 
Therefore, if there are any future directed null geodesics 
existing, which terminate in the past at the singularity, we must have 
$R\rightarrow0$ as $t\rightarrow t_s$ along these curves. Now writing 
\e (\ref{eq:null1}) in terms of variables $(u=r^\alpha,R)$ , we have,
\begin{equation}
\frac{dR}{du}=\frac{1}{\alpha}r^{-(\alpha-1)}R'\left
[1+\frac{\dot{R}}{R'}e^{\psi-\nu}\right]
\label{eq:null2}
\end{equation}
Choosing $\alpha=\frac{5}{3}$ and using \e (\ref{eq:ein4}) 
we get,
\begin{equation}
\frac{dR}{du}=\frac{3}{5}\left(\frac{R}{u}+\frac{\sqrt{v}v'}
{\sqrt{\frac{R}{u}}}\right)\left(\frac{1-\frac{F}{R}}
{\sqrt{G}[\sqrt{G}+\sqrt{H}]}\right)
\label{eq:null3}
\end{equation}

If there are null geodesics which terminate at the singularity 
in the past with a definite tangent, then at the singularity we have 
$\frac{dR}{du}>0$, in the $(u,R)$ plane with a finite value. 
Hence it follows that all points $r>0$ on the singularity curve are 
covered necessarily, because 
$F/R \rightarrow\infty$ with $\frac{dR}{du}\rightarrow-\infty$
for any of these, and hence no outgoing null geodesics can terminate
at these points in the past. 
The central singularity at $r=0$ could however be naked. 
Define the tangent to the outgoing null geodesic from the singularity as,
\begin{equation}
x_0=\lim_{t\rightarrow t_s}\lim_{r\rightarrow 0} \frac{R}{u}
=\left.\frac{dR}{du}\right|_{t\rightarrow t_s;r\rightarrow 0}
\end{equation}
Using \e (\ref{eq:null3}) and (\ref{eq:scurve5}), we then get,
\begin{equation}
x_0^{\frac{3}{2}}=\frac{3}{2}\sqrt{B(0,0)}\X(0)
\label{eq:x01}
\end{equation}

Let us now deduce the necessary and sufficient conditions for
a naked singularity to exist, that is, for null geodesics with a 
well-defined tangent to come out from the central singularity. 
Suppose we have $\X(0)>0$, then we 
always have (from \e (\ref{eq:x01})), $x_0>0$ and 
then in the $(R,u)$ plane, the equation for the null geodesic
that comes out from the singularity is given by 
\begin{equation}
R=x_0u
\label{eq:x02}
\end{equation}
In other words, \e (\ref{eq:x02})
is a solution of the null geodesic equation in the limit of
the central singularity. Thus given $\X(0)>0$, we can always construct
a solution of radially outgoing null
geodesics emerging from the
singularity. This makes the central singularity visible.
In the $(t,r)$ plane, the null geodesics outgoing from  
the singularity will be given as,
\begin{equation}
t-t_s(0)=x_0r^{\frac{5}{3}}
\end{equation}
It follows that $\X(0)>0$ implies $x_0>0$ and we get radially 
outgoing null 
geodesics emerging from the 
singularity, giving rise to the central naked singularity.

On the other hand, if $\X(0)<0$, then we see that the singularity
curve is a decreasing function of $r$. Hence the region around the center
gets singular before the central shell, and the spacetime then
terminates there. In this case, if there were any outgoing 
null geodesic from the central singularity, it must then go to 
a singular region, or outside the spacetime which is impossible. 
Hence when $\X(0)<0$, we always have a black hole solution.

If $\X(0)=0$ then we will have to take into account the next higher 
order non-zero term in the singularity curve equation, and 
do a similar analysis by 
choosing a different value of $\alpha$ in 
equation (\ref{eq:null2}).

We have thus shown above that $\X(0)>0$ is the necessary and
sufficient condition for null geodesics to come out from the central
singularity with a definite positive tangent.
It is thus seen how the initial data,
together with the evolutions chosen in terms of the free functions
such as $b_0$, $\M$ and $h$, 
fully determine 
the final end product
of collapse in terms of either a black hole or a naked singularity.
This is as determined by the $\X(0)$ values above, or more generally 
in terms of the behaviour of the singularity curve in the vicinity of the 
central singularity. This is because $\X(0)$ is determined
by these initial profiles and the evolutions chosen as given by 
\e(\ref{eq:tangent1}), which in turn determine the end states.
Therefore, given any initial regular density and pressure profiles for
the matter cloud from which the collapse develops, there always
exist velocity profiles for collapsing matter shells, and evolutions 
as determined by the Einstein equations,
so that the end state of the collapse would
be BH or NS, depending on the choice made.

As seen above, the different outcomes are characterized by the 
positive or negative values of $\X(0)$.
This in turn depends on the initial values of the functions such as the   
initial mass and velocity profiles, and the evolutions chosen, as
permitted by the Einstein equations. Thus the measures of outcomes
leading to BH/NS phases are accordingly decided (e.g. the entire set of
initial data and evolutions giving $\X(0)>0$ leads to NS).

\section{Conclusions}

While some of the numerical models for perfect fluids indicated
that naked singularities could arise for only a `soft' equation of state,
our results point out that within a generic perfect fluid collapse
scenario with equation of state $p=k\rho$, as such the value of $k$ 
chosen does not appear to have any special significance. 
What matters is the initial data, and the chosen evolutions (i.e. the 
classes of allowed solutions to the Einstein equations), which 
then take this given 
initial data to a specific outcome, depending on the choice made. 
Also, if for a given
chosen evolution, if the value $\X(0)$ was negative (or positive), 
then such will
be the case by continuity for all neighbouring or close by evolutions
where `nearness' is defined in some suitable sense. 
Hence these outcomes in terms of BH or NS may be
considered stable in a certain sense (within spherical symmetry)
as characterized above. 
We thus see that there are classes
of solutions to Einstein equations for perfect fluid models where given
the matter initial data at the initial surface $t=t_i$, these evolutions
take the collapse to end up either as a black hole or the naked 
singularity, depending on the choice of the class. 
This also means that the total space of evolutions can be divided
into distinct subspaces, those that evolve a given initial data 
into black holes, and others that go to a naked singularity.
The results on dust collapse
are of course contained here as a special case with $p=0$.

We should mention here that though the above give some information
on dynamical evolution of collapse while pressures are included in
the analysis, non-spherical perturbations will
have important inputs to decide on the issues such as genericity
and stability. It is possible that methods such as those 
developed in 
[32-33]
could be useful in that direction.

Acknowledgement: It is our pleasure to thank the referees
for their useful comments.

\end{document}